\documentclass[twocolumn,aps,showpacs]{revtex4}
\usepackage{graphicx}
\newcommand{\etal}{{\it et~al.\/}\ }

\newcommand{\aj}{Astron. J.}
\newcommand{\mnras}{Mon. Not. R. Astron. Soc.}
\newcommand{\aap}{Astron. Astrophys.}

\begin{document}
\title{Methods for Constraining Fine Structure Constant Evolution with 
	OH Microwave Transitions}
\author{Jeremy Darling}
%\email{darling@ociw.edu}
\affiliation{Carnegie Observatories, 813 Santa Barbara Street, Pasadena, 
		 CA  91101}
\date{\today}

\begin{abstract}
We investigate the constraints that OH microwave transitions 
in megamasers and molecular absorbers at cosmological distances
may place on the evolution of the fine structure constant
$\alpha=e^2/\,\hbar c$.  The centimeter OH transitions are a 
combination of hyperfine splitting and lambda-doubling 
that can constrain the cosmic evolution of $\alpha$
from a {\it single species}, avoiding systematic errors
in $\alpha$ measurements from multiple species which 
may have relative velocity offsets.
The most promising method compares the 18 and 6 cm OH lines,
includes a calibration of systematic errors, and offers multiple 
determinations of $\alpha$ in a single object.
Comparisons of OH lines to the HI 21 cm line and CO rotational
transitions also show promise.
\end{abstract}
\maketitle

{\it Introduction}---
Recent measurements of the fine structure constant 
$\alpha = e^2/\,\hbar c$ claim a smaller value of $\alpha$ in the past
of order ${\Delta\alpha\over\alpha_\circ}\sim-10^{-5}$ at $z=1$--2 
\citep{mur01,webb01}, but recent 
calculations of the Dirac hydrogen atom spectrum with a dynamic
$\alpha$ reveal new sources of error in the ``many-multiplet'' analysis
of quasar absorption lines \citep{bek03}.
The remaining undisputed measurements are consistent with no evolution 
in $\alpha$ \citep{mur01b,car00}.  Unification
theories that require extra compact dimensions predict variations
in the fundamental constants over cosmic time, including the fine
structure constant (see \citep{mur01} for a review).
New physics is being developed to account for the observed 
properties of the universe, such as the dark energy manifested in the 
cosmological constant, and these theories can be tested by high precision 
observations of the evolution of $\alpha$ over cosmic time.  
Murphy \etal \citep{mur01} have called on the community to produce 
independent measurements of the fine structure constant evolution to 
verify current results from quasar absorption lines.  
We present a method to exploit the hyperfine 
structure of the OH molecule 
to bypass the systematic pitfalls of other radio and 
sub-millimeter determinations of $\alpha(z)$.  

Centimeter OH transitions can be observed at cosmological distances
in absorption against strong radio continuum sources or in OH megamasers
(OHMs) \citep{kan02,dar02b}.  OHMs are luminous natural 
masers found in the nuclei of major galaxy mergers \citep{dar02a}.
They are detectable at high redshift, 
and deep surveys at numerous radio telescope facilities are expected
to identify many OHMs at medium and high redshifts \citep{bri98,dar02b}.
OHMs are luminous, can be observed with high spectral 
resolution, and like OH absorbers can be narrow and spatially compact, 
marking with high precision a specific position and redshift which reduces
potential systematic errors.  
The frequency shift in the main 18 cm OH lines due to a change in $\alpha$ is 
considerable:  $\sim30$ kHz for ${\Delta\alpha\over\alpha_\circ} = 10^{-5}$.
Resolution of such a shift is trivial for typical observations of
OH lines; the main difficulty lies in identifying the true redshift
of a given galaxy because all lines may be influenced by a changing 
fine structure constant.  
An ideal measure of $\Delta\alpha\over\alpha_\circ$ would be obtained from 
ratios of lines that are spatially coincident with identical velocity 
structure.  The multiple microwave transitions 
in the OH molecule may provide such an ideal diagnostic of the evolution of
$\alpha$ provided that redshifts can be measured to at least one part in
$10^5$.  

{\it The OH Molecule}---
Each rotation state of the OH molecule is split by lambda-type doubling
--- the interaction of electronic angular momentum with the molecular
rotation --- and each of these states is further split by hyperfine 
splitting \citep{gor70,dou55}.  
The net result is that each of the astronomically observed 
microwave transitions of OH is both a hyperfine transition and a 
transition between lambda-doubled levels.  The two splittings 
depend differently on the fine structure constant $\alpha$.  
Hyperfine transition frequencies in OH depend on terms of order
$\mu_\circ \mu_I/(I\,\hbar\,a_\circ^3)$ where $\mu_\circ=e\,\hbar/2m_ec$
is the Bohr magneton, $\mu_I$ is the nuclear magnetic moment
($\mu_I\propto\mu_\circ$), $I$ is the nuclear spin, and 
$a_\circ=\hbar^2/m_e e^2$ is the Bohr radius \citep{dou55}.
Hyperfine transition frequencies in OH thus follow the same $\alpha$
dependence as the HI 21 cm transition:  $\nu_{HF}\propto \alpha^4$.

The lambda doubling in OH depends on the molecular
state.  For the $^2\Pi_{3/2}$ state, which includes the OH ground
state, the leading term in the lambda doubling energy is independent of
$\alpha$:
$B^3/(A\,E_{\Sigma-\Pi})\propto\alpha^0$
where $B\propto\alpha^2$ is the rotational constant, $A\propto\alpha^4$
is the spin-orbit coupling constant (also called the fine structure 
interaction constant), and $E_{\Sigma-\Pi}\propto\alpha^2$ is the 
energy between the $\Sigma$ and $\Pi$ electronic states 
\citep{van29,dou55,gor70,bro79}.  
The $A$, $B$, and $E_{\Sigma-\Pi}$ terms also depend on 
the fundamental constants $m_e$, $c$, and $\hbar$.  For the $^2\Pi_{1/2}$
state, the dominant term in the lambda doubling energy does depend on 
$\alpha$:  $A\,B/E_{\Sigma-\Pi}\propto\alpha^4$ \citep{van29,bro79}.
Second order corrections modify the $\alpha$ dependence 
of $A\propto\alpha^4$ at the 10--$25\%$ level:  $\nu_{3/2}\propto\alpha^{0.4}$
and $\nu_{1/2}\propto\alpha^{5.0}$.  Higher order corrections modify
the corrected exponent by $\lesssim5\%$ \citep{mul31,des77a,bro79}, and these 
corrections are only valid for ${\Delta\alpha\over\alpha_\circ}\ll 1$.
%, which is ultimately a multiplicative 
%factor relating $\Delta\alpha\over\alpha_\circ$ to a redshift difference
%between lines $\Delta z$.

Hence, a generic $^2\Pi_{3/2}$ OH microwave transition (ignoring pure
hyperfine transitions) can be written 
in terms of the fine structure constant as
$\nu_{3/2} = \Lambda\alpha^{0.4} \pm (\Delta^+ \pm \Delta^-)\alpha^4$ 
where the choices of
sign are independent (there are 4 possible lines), $\Lambda$ sets the
frequency of the lambda doubled splitting, and $\Delta^\pm$ sets the 
strength of the hyperfine splitting for the $\pm$ parity lambda state.
A generic $^2\Pi_{1/2}$ state can likewise be expressed as 
$\nu_{1/2} = \Lambda\alpha^5 \pm (\Delta^+ \pm \Delta^-)\alpha^4$ where the 
constants have different numerical values from the $^2\Pi_{3/2}$ case.
Comparison of pairs of microwave transitions can thus 
determine the value of $\alpha$ in cosmic OH masers; one line fixes the
velocity of the source, and the other determines $\alpha$.  The remarkable
properties of the OH molecule provide a means to measure or constrain 
possible cosmic evolution of $\alpha$ {\it from a 
single species with built-in checks on systematic errors}.  Previous 
measurements of $\alpha$ are susceptible to systematics
such as velocity offsets between species which can mimic a changing 
$\alpha$ \citep{car00,mur01b}.
As illustrated in detail below, intercomparison of OH lines provides cases
with no dependence on $\alpha$ which serve as benchmarks to quantify the
systematics present in cases with a strong dependence on $\alpha$.  This
property of the OH microwave transitions may provide the highly constrained
method required to investigate claims of a time-varying fine structure  
constant.  
\begin{figure}
\includegraphics[scale=0.6]{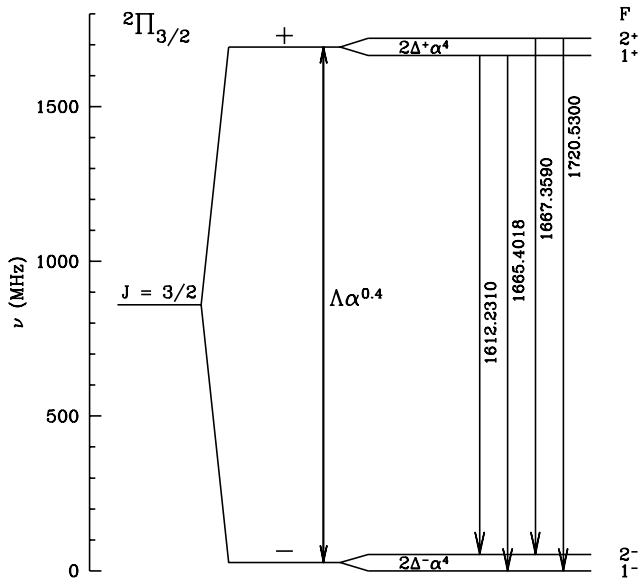}
\caption{The lambda doubling and hyperfine splitting of the $^2\Pi_{3/2}$ 
ground state of the OH molecule to scale.  The 18 cm transitions are 
as labeled in MHz, and the parameters $\Lambda$ and $\Delta^\pm$ are 
related to the 
size of the lambda doubling and hyperfine splitting, respectively.
The zero point of the diagram is arbitrarily set to the lowest energy level.
\label{fig}}
\end{figure}

%\section{Measuring $\alpha$ Strictly from OH Transitions}

{\it OH Ground State Transitions}---
The four 18 cm OH transitions in the ground $^2\Pi_{3/2}\ J=3/2$ 
rotation state of
OH are a combination of lambda type doubling and hyperfine splitting
\citep{dou55} (Fig. 1).  The
dominant contribution comes from lambda doubling of order 1666 MHz, and each
of these is doubled by hyperfine splitting of order 54 MHz.  The hyperfine
splitting is unequal between the upper and lower lambda doubled states by 
2 MHz, so that four 18 cm lines are possible in the ground rotation state
of OH rather than two degenerate lines \citep{des77a,des77b}:
\begin{eqnarray}
  \nu_{1612} &=& \Lambda\alpha^{0.4} - (\Delta^++\Delta^-)\alpha^4\\
  \nu_{1665} &=& \Lambda\alpha^{0.4} - (\Delta^+-\Delta^-)\alpha^4\\
  \nu_{1667} &=& \Lambda\alpha^{0.4} + (\Delta^+-\Delta^-)\alpha^4\\
  \nu_{1720} &=& \Lambda\alpha^{0.4} + (\Delta^++\Delta^-)\alpha^4
\end{eqnarray}
where $\Lambda$ sets the magnitude of the lambda doubling and $\Delta^+$ 
and $\Delta^-$
set the magnitude of the hyperfine splitting of the upper and lower 
lambda-doubled states, respectively (Fig.\@ \ref{fig}).
For $\alpha_\circ=0.007297352533(27)$ 
(1998 CODATA recommended value), 
$\nu_{1665} = 1665.40184(10)$ MHz, and
$\nu_{1667} = 1667.35903(10)$ MHz \citep{ter72}, 
we obtain $\Lambda = 11926.36309(51)$ MHz. 
For $\nu_{1720} = 1720.52998(10)$ MHz, 
$\Delta^+ = 9.720353(25)\times10^9$ MHz, and
$\Delta^- = 9.375256(25)\times10^9$ MHz.
These values for $\Lambda$ and $\Delta^\pm$ determine a frequency for 
the final line
of $\nu_{1612} = 1612.23089(12)$ MHz, which agrees with the measured value
of 1612.23101(20) MHz \citep{ter72}.  
From $\alpha_\circ$ ($\alpha$ today) and the derived coefficients, we 
obtain the size of the lambda doubling and the hyperfine splitting:
$\nu_\Lambda = \Lambda\alpha_\circ^{0.4} = 1666.38044(7)$ MHz, 
$2\Delta^+\alpha_\circ^4 = 55.12814(14)$ MHz, and
$2\Delta^-\alpha_\circ^4 = 53.17095(14)$ MHz.
Note that $\nu_\Lambda$ has a value equal to the mean of the 
1667 and 1665 MHz lines and the mean of the 1720 and 1612 MHz lines;
this is the closure criterion for the ground rotation state of OH.  

The main lines observed in OH megamasers are at 1665 and 1667 MHz, with 
the latter dominant.  We can form two quantities from the observed 
frequencies of these lines which isolate powers of $\alpha$:
\begin{eqnarray}
  \Delta\nu &\equiv& \nu_{1667}-\nu_{1665} = 2(\Delta^+ - \Delta^-) \alpha^4\\
  \Sigma\nu &\equiv& \nu_{1667}+\nu_{1665} = 2\Lambda\alpha^{0.4}
\end{eqnarray}
such that the 
ratio $Y=\Delta\nu/\Sigma\nu=\alpha^{3.6}(\Delta^+-\Delta^-)/\Lambda$.
In terms of the fractional change in this ratio at a redshift $z$ and the
difference in redshift derived from the separation and average of
the 1665 and 1667 MHz lines, 
\begin{equation}
  {\Delta Y\over Y}\equiv{Y_{z}-Y_{\circ}\over Y_{\circ}}
  	= {z_{\Sigma\nu} - z_{\Delta\nu}\over 1+z_{\Delta\nu}}
       = {\alpha^{3.6} - \alpha_\circ^{3.6}\over\alpha_\circ^{3.6}}
%	= \left(\alpha\over\ \alpha_\circ\right)^{3.6} - 1
		    \simeq 3.6\,{\Delta\alpha\over\ \alpha_\circ}
\end{equation}
where
\begin{equation}
  {\Delta\nu_\circ\over\Delta\nu_z} \equiv 1 + z_{\Delta\nu},\ \ \ \ 
  {\Sigma\nu_\circ\over\Sigma\nu_z} \equiv 1 + z_{\Sigma\nu},
    							 \label{eqn:zdefs}
\end{equation}
$\Delta\alpha = \alpha - \alpha_\circ$, and 
$|\Delta\alpha|\ll\alpha_\circ$.  The difference between the
measured redshifts of the difference and sum of the main OH lines is
thus of the same order of magnitude as the fractional change in the fine
structure constant.  
Note that the change in $\Delta\nu$ expected for 
${\Delta\alpha\over\alpha_\circ} = 10^{-5}$ is of order 100 Hz, which is 
the accuracy to which the 18 cm OH line frequencies have been measured 
experimentally.  Better determinations of the line frequencies are required.
Note, however, that a $\Delta\nu$ formed from the 1720 and 1612 MHz lines
would have the same dependence on $\alpha$ but have substantially lower
spectral resolution requirements than above (the change is of order 4 kHz, 
a factor of 55 larger).  The 
OH satellite lines at 18 cm have been detected in only a few local starburst
galaxies \citep{sea97,fra98},
so the prospects for detecting these lines at higher redshifts are poor.
Detection of even a single satellite 18 cm line would provide a
useful constraint on $\alpha$ (a factor of 27 better than is possible with 
$\Delta\nu$).

The evolution in the fine structure constant can be obtained from comparisons
of any two OH lines if they originate from the same physical region.  The
most likely lines are the main 18 cm OH lines at 1667 and 1665 MHz:
\begin{equation}
	{z_{1665} - z_{1667} \over 1+z_{1667}}
		\simeq 1.8 \left(\Delta\alpha\over\alpha_\circ\right)\,
		\left(\Sigma\nu\ \Delta\nu\over
		  \nu_{1667}\ \nu_{1665}\right)_\circ    
  \label{eqn:linediff}
\end{equation}
The constant factor on the right hand side of Eqn. \ref{eqn:linediff}
refers to the rest frequencies of the OH lines and has a value of 0.00235.
This is a poor method for constraining fine structure constant evolution 
because it requires extremely precise redshift measurements.  The comparison
of the 1667 and 1665 MHz lines combine like powers of $\alpha$, so the 
size of the effect reduces to the ratio between the difference in hyperfine
splittings ($\Delta\nu \simeq 2$ MHz) and the line frequency.  
More sensitive measurements should use ratios of lines with different
dependence on $\alpha$ such as the 5 cm transitions of OH.
Regions with 5 or 6 cm OH transitions are likely to be 
physically conterminous with 18 cm and CO transition regions 
\citep{hen87,cas01}, especially in absorption systems.

%\subsection{18 cm OH vs 5 and 6 cm OH}

{\it 6 cm transitions ($^2\Pi_{1/2}\ J=1/2$)}---
The OH 6 cm transitions have been detected in absorption in five
nearby OHMs, and the properties of the 18 cm and 6 cm lines appear
to be correlated \citep{hen87}.
The three 6 cm $^2\Pi_{1/2}\ J=1/2$ OH lines have frequencies 
4660.242(3), 4750.656(3), and 4765.562(3) MHz \citep{rad68}.
% (is $0^-\rightarrow0^+$ forbidden?)
These can be expressed in a similar manner to the 18 cm transitions:
\begin{eqnarray}
  \nu_{4660} &=& \Lambda_6\alpha^5 - (\Delta_6^-+\Delta_6^+)\alpha^4\\
  \nu_{4751} &=& \Lambda_6\alpha^5 + (\Delta_6^--\Delta_6^+)\alpha^4\\
  \nu_{4766} &=& \Lambda_6\alpha^5 + (\Delta_6^-+\Delta_6^+)\alpha^4
\end{eqnarray}
Note that the ``$+$'' and ``$-$'' states in $^2\Pi_{1/2}\ J=1/2$ are
reversed from the $^2\Pi_{3/2}\ J=3/2$ ground state; the 4751 MHz line
is the analog to the 1667 MHz line \citep{des77b}.  From the laboratory values
for the line frequencies, we obtain values for the 6 cm coefficients:
$\Lambda_6 = 2.2775178(10)\times10^{14}$ MHz, 
$\Delta_6^- = 1.59421(7)\times10^{10}$ MHz, 
and $\Delta_6^+ = 2.6283(7)\times10^9$ MHz.  Hence, the hyperfine splittings
are quite unequal:   
$2\Delta_6^+\alpha^4_\circ = 14.906(4)$ MHz, and 
$2\Delta_6^-\alpha^4_\circ = 90.414(4)$ MHz.  
LTE ratios of the 4660, 4751, and 4766 MHz lines are 1:2:1 \citep{JPL}.  
Observations of the nearest OHMs find that the absorption in these lines 
deviates somewhat from LTE, but the 4751 MHz line still tends to dominate
\citep{hen87}.  

Since the 6 cm lines are of nearly equal strength, it is likely that
if any are detected, there will be at least two detectable lines 
\citep{hen87}.  Comparison of well-separated pairs of 6 cm lines produces 
$\Delta\alpha\over\alpha_\circ$ ``gain'' factors of order 
$\Delta\nu_6/\nu\simeq0.02$ where $\Delta\nu_6$ is the line separation.
Comparing the dominant lines at 18 and 6 cm we obtain
\begin{equation}
	{z_{4751}-z_{1667}\over 1+z_{1667}} \simeq 
	  -{\Delta\alpha\over\alpha_\circ}
	  \left(1.8\,{\Sigma\nu\over\nu_{1667}} 
	+ {1\over2}\,{\Sigma\nu_6\over\nu_{4751}}\right)_\circ \label{eqn:4751}
\end{equation}
where $\Sigma\nu_6=2\Lambda_6\alpha_\circ^5$ and the constant term is
4.59.
This is a dramatic improvement over the 1665 to 
1667 MHz line comparison (by a factor of nearly 2000), and 
for $\Delta\alpha/\alpha_\circ = 10^{-5}$, the resolution required for
OH lines is 10's of kHz which is easily achieved.  
Equation \ref{eqn:4751} applies to the
comparison of {\it any} 18 cm line to {\it any} 6 cm line, offering
the possibility for multiple measurements of $\alpha$ from a single
system.  
%One can identify systematics and the true size of measurement
%uncertainties by comparing redshifts of the different 6 cm lines because 
%they all depend on $\alpha$ in an identical manner.  
One can also obtain an accurate 
$\alpha$-independent zero point by comparing $\Delta\nu$ to 
$\Delta\nu_6=2(\Delta_6^-+\Delta_6^+)\alpha^4$
to reveal any velocity offsets between 18 cm and 6 cm OH regions:
\begin{equation}
{z_{\Delta\nu_6}- z_{\Delta\nu} \over 1+z_{\Delta\nu}} =0
\end{equation}
Hence, comparison of 18 and 6 cm OH lines offers a sensitive method for
detecting changes in $\alpha$ that includes redundant checks on statistical
and systematic errors.

{\it 5 cm transitions ($^2\Pi_{3/2}\ J=5/2$)} ---
Absorption in a 5 cm OH line has been detected in just one OHM, 
Arp 220 \citep{hen86}.
The 5 cm OH lines have frequencies 
6016.746(5), 6030.7485(2), 6035.0932(2), and 6049.084(8) MHz
\citep{rad68,mee75,JPL}.
These can be expressed in a similar manner to the 18 cm transitions:
\begin{eqnarray}
  \nu_{6017} &=& \Lambda_5\alpha^{0.4} - (\Delta_5^- + \Delta_5^+) \alpha^4\\
  \nu_{6031} &=& \Lambda_5\alpha^{0.4} - (\Delta_5^- - \Delta_5^+) \alpha^4\\
  \nu_{6035} &=& \Lambda_5\alpha^{0.4} + (\Delta_5^- - \Delta_5^+) \alpha^4\\
  \nu_{6049} &=& \Lambda_5\alpha^{0.4} + (\Delta_5^- + \Delta_5^+) \alpha^4
\end{eqnarray}
Note that the ``$+$'' and ``$-$'' states in $^2\Pi_{3/2}\ J=5/2$ are
reversed from the $^2\Pi_{3/2}\ J=3/2$ ground state; the 6035 MHz line
is the analog to the 1667 MHz line \citep{des77b}.  From the laboratory values
for the first three line frequencies, we obtain values for the 5 cm 
coefficients:
$\Lambda_5 = 43177.898(1)$ MHz, 
    $\Delta_5^- = 3.2350(9)\times10^9$ MHz, 
and $\Delta_5^+ = 2.4690(9)\times10^9$ MHz.
From these, we predict $\nu_{6049} = 6049.096(4)$ MHz, which is 
in fair agreement with the measured value.
The hyperfine splittings in this case are only slightly unequal:  
$2\Delta_5^+\alpha^4_\circ = 14.0025(50)$ MHz, and 
$2\Delta_5^-\alpha^4_\circ = 18.3472(50)$ MHz.  
LTE ratios of the 6017, 6031, 6035, and 6049 MHz lines are 1:14:20:1 
\citep{JPL}.

Comparing the dominant lines at 18 and 5 cm we obtain
\begin{equation}
	{z_{6035}-z_{1667}\over 1+z_{1667}} \simeq 
	  1.8\,{\Delta\alpha\over\alpha_\circ}\,
	  \left(-{\Sigma\nu\over \nu_{1667}} +
		{\Sigma\nu_5\over \nu_{6035}}\right)_\circ
\end{equation}
where $\Sigma\nu_5 = 2\Lambda_5\alpha^{0.4}$. 
The constant term is 0.00045, which is smaller than the 1665 to 
1667 MHz line comparison by a factor of 5.  Comparing any 
18 cm line to any 5 cm line gives the same order of magnitude $\alpha$
``gain'' factor, to within a factor of 2.  
The difference in the hyperfine splitting between the lambda-doubled 
levels of $^2\Pi_{3/2}\ J=5/2$ is so small that it provides poor 
leverage on $\alpha$ and requires extremely accurate redshift determinations.
The hyperfine splitting of these levels is small overall, so detection
of the satellite 5 cm lines would be of limited use (and unlikely).  

%\section{Measuring $\alpha$ from Combinations of OH and Other Species}
{\it OH vs HI}---
The 21 cm hyperfine transition of HI is proportional to 
$\mu_p\mu_\circ/(\hbar a_\circ^3)$ 
where $\mu_p = g_pe\ \hbar/(4m_pc)$ and $g_p$ is the proton $g$-factor 
\citep{mur01b}.  In terms of $\alpha$, the 21 cm line frequency 
is proportional to $\alpha^4(g_pm_e^2/m_p)$ modulo factors
of $\ \hbar$ and $c$.  The ratios of the 1667 or 1665 MHz lines or their sum
to the 21 cm line can thus provide a measurement of $\alpha$:
\begin{equation}
 {z_{HI}- z_{1667} \over 1+z_{1667}}
		\simeq -1.8\,{\Delta\alpha\over\alpha_\circ}
		\left({\Sigma\nu\over \nu_{1667}}\right)_\circ
		\simeq -3.6\,{\Delta\alpha\over\alpha_\circ}
\end{equation}
Comparison of the 1665 MHz line or $\Sigma\nu$ to HI produces the same 
relationship.
For $\Delta\alpha/\alpha_\circ = 10^{-5}$, redshifts must be determined
to about 4 parts in $10^5$.  This method does not require detection of
the 1665 MHz line, but if it is detected, it provides a second determination
of $\alpha$.  This is a promising avenue to 
measure $\alpha(z)$, with a check of systematics provided by
the $\alpha$-independent ratio $\Delta\nu/\nu_{HI}$:
\begin{equation}
 {z_{HI}- z_{\Delta\nu} \over 1+z_{\Delta\nu}} = 0
\end{equation}
The $\Delta\nu/\nu_{HI}$ ratio can provide an anchor
for the method and indicate the influence of physical and/or velocity 
offsets between OH and HI, whereas the $\nu_{1667}/\nu_{HI}$ ratio provides
the maximum detectability of a change in $\alpha$.  

{\it OH vs CO}---
The rotational transitions of CO (and other diatomic molecules with balanced 
electronic angular momentum)
have frequencies proportional to $\ \hbar/(M a_\circ^2)$ where $M$ is 
the reduced mass \citep{gor70}.  
In terms of $\alpha$, the rotational
transitions of CO are proportional to $\alpha^2 m_e^2/M$ modulo factors
of $\ \hbar$ and $c$.  The ratio of the 1667 or 1665 MHz line to a CO 
rotational transition can thus provide a measurement of $\alpha$:
\begin{equation}
  {z_{CO}- z_{1667} \over 1+z_{1667}}
		\simeq -1.6\,{\Delta\alpha\over\alpha_\circ}\,
		\left(\nu_{1665}\over \nu_{1667}\right)_\circ  
\end{equation}
For the 1665 MHz line, the constant term is inverted and differs from
unity by about 0.1$\%$.
For ${\Delta\alpha\over\alpha_\circ} = 10^{-5}$, redshifts must be determined
to about 2 parts in $10^5$.  While comparisons of CO to OH transitions do
not offer any $\alpha$-independent line ratios, the sum and difference
of 18 cm OH lines do offer some leverage on systematic errors:
\begin{equation}
  {z_{CO}- z_{\Delta\nu} \over 1+z_{\Delta\nu}}
		\simeq 2\,{\Delta\alpha\over\alpha_\circ}\ ;\ 
  {z_{CO}- z_{\Sigma\nu} \over 1+z_{\Sigma\nu}}
		\simeq -1.6\,{\Delta\alpha\over\alpha_\circ}
\end{equation}

{\it Conclusions}---
The remarkable properties of the microwave transitions in the OH molecule 
provide a robust method to measure deviations in $\alpha$ over cosmic time 
%without reference to other species.  This bypasses the difficult 
%problem of identifying spatially coincident regions which emit or absorb 
%in more than one reliable narrow line such as CO or HI.  
%(While the 1665 MHz line is believed to be more extended than the 1667
%MHz line, the two lines in OH megamasers are likely to be 
%spatially coincident in aggregate (ie --- spatially unresolved spectra))
%\citep{pih01,rov03}, reducing susceptibility to systematic errors 
%endemic to studies requiring the comparison of multiple species.
%OH absorbers...HF ratio consistent with TDE
{\it from a single species}.  This 
approach eliminates the largest systematic errors present in other 
determinations of $\alpha$ and provides estimates of the remaining
statistical and systematic errors.  
The most promising method for measuring $\alpha$ is the comparison of 18
and 6 cm OH lines.  This method includes $\alpha$-independent line ratios
which can identify the true size of statistical and systematic errors.  Also
promising are comparisons of OH lines to the HI 21 cm line and CO and
other molecular rotation transitions, but only HI provides checks on 
systematics.

Deep surveys for OH megamasers (and OH gigamasers) are underway from 
the local universe to $z\simeq4$ and a subset of the new discoveries will
have spectra appropriate for measurements of $\alpha$.  
Several OH absorption systems have already been identified out to $z=0.9$
\citep{kan02} and more will be found in the near future.
In the meantime, 
more precise laboratory measurements of the microwave transitions in OH 
would eliminate some of the uncertainty in the proposed techniques.  

%\begin{acknowledgments}
The author thanks the anonymous referees for insightful and 
thoughtful comments which significantly improved the content 
of this presentation.
It is a pleasure to thank John Brown for critical discussions
and John Grula for library assistance.  
%\end{acknowledgments}

\end{document}